\documentclass[twocolumn]{aastex61}
\usepackage{longtable}
\usepackage{graphicx}

\received{...}
\revised{...}
\accepted{\today}
\submitjournal{ApJ}

\shorttitle{XRBs in the low-luminosity regime}
\shortauthors{Sonbas et al.}

\begin{document}
\title{A Study of Low-Mass X-Ray Binaries in the Low-Luminosity Regime}
\email{edasonbas@gmail.com}

\author{E. Sonbas}
\altaffiliation{Department of Physics, The George Washington University, Washington, DC 20052, USA}
\affil{University of Adiyaman, Department of Physics, 02040 Adiyaman, Turkey}

\author{K. S. Dhuga}
\affil{Department of Physics, The George Washington University, Washington, DC 20052, USA}

\author{E. G\"o\u{g}\"u\c{s}}
\affil{Sabanc\i~University, Orhanl\i~- Tuzla, Istanbul 34956, Turkey}



\begin{abstract}
A recent study of a small sample of X-ray binaries (XRBs) suggests a significant softening of spectra of neutron star (NS) binaries as compared to black hole (BH) binaries in the luminosity range 10$^{34}$  -  10$^{37}$ erg/s. This softening is quantified as an anticorrelation between the spectral index and the 0.5 - 10 keV X-ray luminosity. We extend the study to significantly lower luminosities  (i.e., $\sim$ a few $\times$  $10^{30}$ erg/s) for a larger sample of XRBs. We find evidence for a significant anticorrelation between the spectral index and the luminosity for a group of NS binaries in the luminosity range 10$^{32}$ to 10$^{33}$ erg/s. Our analysis suggests a steep slope for the correlation i.e., -2.12 $\pm$ 0.63. In contrast, BH binaries do not exhibit the same behavior. We examine the possible dichotomy between NS and BH binaries in terms of a Comptonization model that assumes a feedback mechanism between an optically thin hot corona and an optically thick cool source of soft photons. We gauge the NS-BH dichotomy by comparing the extracted corona temperatures, Compton-y parameters and the Comptonization amplification factors: The mean temperature of the NS group is found to be significantly lower than the equivalent temperature for the BH group. The extracted Compton-y parameters and the amplification factors follow the theoretically predicted relation with the spectral index.

\end{abstract}




\section{Introduction} \label{sec:intro}
Studies of Low-Mass X-ray binaries (LMXBs, both transient and quasi-persistent) show that their spectra, in the energy range $\sim$0.5 - 10.0 keV, exhibit features that typically require at least two components for a statistically acceptable description: a thermal and a non-thermal component. The thermal component is usually modeled as a single or a multicolor blackbody \citep{2009ApJ...696.1257L} and is thought to be associated with the inner parts of a geometrically thin but optically thick accretion disk. In the case of NS binaries, the surface of the NS is also potentially a significant source of thermal emission. On the other hand, there is considerable debate regarding the origin of the non-thermal component, which is typically represented as a power-law (PL) depicting the photon spectrum \textit{N(E)} as \textit{E$^{-\Gamma}$}, where $\Gamma$, the photon index of the power-law typifies whether the spectrum is soft ($\Gamma$ $>$ 2) or hard ($\Gamma$ $<$ 2). A model that is widely discussed in the literature \citep{2000A&A...361..175M, 2002ApJ...572L.173L} posits the existence of a hot corona of electrons in the vicinity of the compact object that acts as a diffuse and energetic source for the scattering of soft photons. The nature, and indeed, the geometrical profile of the corona, including the origin of the soft seed photons that is a critical component of the scattering process, are largely unknown and are of considerable theoretical interest, particularly in our understanding of the various spectral states that BH and NS binaries are known to exhibit\citep{2006ARA&A..44}.\\
\\
A number of spectral states and associated transitions, which are usually depicted as tracks on a hardness-intensity diagram (HID; Belloni 2010 and references therein), are discussed in the literature (Remillard \& McClintock, 2006, Belloni, 2010), however, the two most prominent states, of which there is universal agreement, are the Low/Hard (LHS) and the High/Soft states (HSS). Other states such as the Hard/Intermediate state (HIMS), Soft/Intermediate state (SIMS) are less certain (Done et. al. 2007, Homan and Belloni 2005, Dunn et. al. 2010). The quiescent state, which often gets linked together with the LHS rounds out the categories (Belloni 2010). Other definitions have been proposed: one such set (Remillard and McClintock 2006) contains three spectral states: hard, thermal and the steep power law (SPL). The HID represents an attempt to correlate the important global properties such as the accretion rate, physical size of the accretion disk and the prime emission processes that are thought to be the main underlying drivers of the observed states and the transitions among them. Based on spectral and timing properties, the hard and the thermal states are readily identified with the LHS and HSS respectively. However, the SPL state is not so easily characterized; it apparently encompasses or overlaps the two intermediate states: HIMS and SIMS.  This is potentially a source of confusion and uncertainty especially when comparing spectral properties of a sample of LMXBs based solely on a luminosity scale.\\
\\
\citet{2015MNRAS.454.1371W} point out that the photon index of a simple PL that provides an adequate description of the spectra for a small sample of LMXBs, increases sharply with decreasing luminosity in the range $10^{34} - 10^{36}$ erg/s. In other words, the spectra of these particular (NS) LMXBs soften with decreasing luminosity. This phenomena has been observed before \citep{2011MNRAS.417..659A, 2014ApJ...780..127B, 2014MNRAS.438..251L, 2013ApJ...773...59P} but mostly for individual sources: \citet{2015MNRAS.454.1371W} on the other hand assert that this feature may be universal, therefore present in a larger group of NS binaries that fall into the low-accretion, low-luminosity category. For BH systems this behavior is thought to be a symptom of radiative inefficiency for low-accretion flow \citep{2013ApJ...773...59P}. \citet{2015MNRAS.454.1371W} compare the $\Gamma$-Luminosity correlation seen in NS binaries with a small sample of BH binaries and note that the BH binaries in the same low-luminosity range have significantly harder spectra thus leading to the speculation that this different spectral behavior is a possible signature that could be used to distinguish between BH and NS binaries. 
It is noted that the XRB spectra in or near a threshold luminosity of $\sim 10^{34}$ erg/s, exhibit very distinct features that are either predominantly PL-like or thermal-like or in some cases, the result of an admixture of the two components. High quality data for NS systems however seem to require the presence of an additional soft component that is possibly linked to the NS surface \citep{2013ApJ...767L..31D}. Moreover, very recent studies \citep{2013ApJ...774..131C, 2011ApJ...736..162F, 2014ApJ...791...47D, 2014ApJ...795..131H, 2016ApJ...833..186M} of quasi-persistent NS binaries in or near the quiescent state exhibit a strong thermal component as well. Thus presently it is not entirely clear whether the softening in the NS binaries is due to the increasing influence of the thermal component, likely from the inner parts of the disk or from the NS surface itself, or due to some, as yet unidentified, underlying PL component that evolves as a function of luminosity. \\
\\
In this work, we report on the spectral analysis of 27 XRBs. We present our results in the context of the reported correlation between the PL index and the 0.5-10 keV X-ray luminosity and its possible interpretation in terms of Comptonization of seed photons via the hot corona in the vicinity of the central compact object. The paper is organized as follows: In section 2, we provide a more detailed scope of the work. In section 3 we report the details of the source selection criteria, data reduction and methodology. In section 4, we present the main results of our analysis based on spectral fits using PL + blackbody models, as well as, the more physically motivated Comptonization model COMPPS. We conclude by providing a summary of the main findings in section 5.
\section{Aims and scope of the work}
In this section, we begin by recalling the most pertinent facts relevant to our study. Wijnands et al. (2015) claim that a simple PL, with a relatively large spectral index, provides a satisfactory description of the 0.5 - 10 keV spectra of a number of NS binaries in the low-luminosity regime. Further, the authors note that the spectra of these particular NS binaries soften quite sharply with decreasing luminosity. On the other hand, comparable BH binary spectra do not exhibit the same softening. The `divergence' effect between the NS-BH binaries appears to occur at or near a threshold luminosity of a few $\times$ 10$^{34}$ erg/s, a luminosity that represents a very small fraction ($\sim$ 0.0001) of Eddington, one that is typically associated with LH/quiescent states. 
Armas Padilla et. al (2013), in a study of three persistent but very faint NS-binaries, also find that a thermal component is not required for one of their sources; apparently a PL, again with a relatively large spectral index, is all that is needed. Indeed, Armas Padilla et. al. (2013) question why the soft component is not detected in the source, and note that it is unlikely to be related to the quality of the data because all three sources were well measured. Furthermore, they performed a series of simulations to gauge the effect of absorption and concluded that they would have detected the thermal component in their source had it been present at a significant level. This leads them to the intriguing speculation that the spectral difference must be intrinsic to the source and that its unusually soft spectrum (large $\Gamma$ ) is a feature not related to a thermal component (which presumably would be due to the NS surface or the accretion disk) but instead is potentially tied to the underlying PL itself.\\ 
\\
A physical process that naturally produces a PL-like behavior is Comptonization; this is an efficient cooling mechanism whereby energetic electrons impart energy through collisions with low energy photons. Thus the emergent spectrum from the medium is critically dependent on the supply and the temperature of these seed photons. The origin of these seed photon is commonly assumed to be either the accretion disk or the NS surface. Regardless of the origin, photon scattering in the vicinity of a hot corona of electrons is complex in that it is necessary to properly account for photon feedback, where upscattered photons are subsequently absorbed by the source of the photons and thereby raising the local temperature and thus increasing the seed photon population which in turn can expedite the Comptonization process. The overall effect is further cooling and hence continual softening of the emerging spectrum. A natural question then, is Comptonization a sufficiently significant component in the low-accretion, low-luminosity regime, particularly for those spectral states that fall into the (somewhat ambiguous) SPL category or is there another process that produces a steep PL-like  spectral behavior that is apparently observed in the few transients measured thus far? \\
\\
The aims of our study are as follows: a) to probe the robustness of the $\Gamma$ - L$_{x}$ (anti)correlation through the spectral analysis of a number of faint LMXB sources (both transient and quasi-persistent) and determine whether the steep softening of the spectra occurs over a large dynamic range in luminosity, b) to compare NS and BH binaries in order to probe possible dichotomies between the two groups, and c) to test whether the SPL component in the spectra is adequately described by Comptonization or whether an additional PL-like component is needed. Comparitively, our work can be seen as an extension of the study of Wijnands et al. (2015) to lower luminosities and a larger sample set. In addition, we deploy the Comptonization model to interpret some of our results. Our data sets are selected from the XMM-Newton and Chandra archives and we limit these data primarily to low-accretion, low-luminosity LMXBs. We do not include accreting millisecond pulsars (AMXPs) in our target list. Emission from these sources arises primarily through a) accretion via magnetic poles whereby significant magnetic fields create columns of inflow and outflow, leading to regions where shocks form, and b) through non-accretion mechanisms involving the collision of the pulsar wind with the materials ejected from the donor; both of these mechanisms tend to produce hard spectra. Our targets of study are accreting sources with low magnetic fields and thus form a different group altogether. Effort is made to achieve acceptable fits with use of single models i.e., thermal, phenomenological PL, and Comptonization. As needed, minimum combination of models are used in subsequent fits. The aim is to extract uniform global trends of the fitted parameters (for both NS and BH binaries) as opposed to fine-tuning the models for best statistical outcomes. The thermal component is modeled by a multicolor accretion disk or a standard blackbody; in addition to the ordinary PL, a PL-like component can be realized via Comptonization models, such as COMPPS. Knowing that Comptonization tails typically show up at high energies, we adopt some of the parameterizations of Burke et al. (2017) who studied Comptonization in NS-BH binary systems in the RXTE energy range. For example, the electron distribution is assumed to be Maxwellian and the geometry is taken to be spherical. The corona electron temperature and the optical depth are treated as free parameters. The known galactic absorption is included explicitly and an intrinsic absorption is left as a free parameter. This places us in a position to explore possible correlations among the main parameters i.e., $\Gamma$, luminosity, and the Compton y-parameter, which quantifies the average change in energy of photons as they traverse the hot corona of electrons depicted by an electron cloud temperature kT$_{e}$ and optical depth $\tau$.\\

\section{Sample selection and methodology}
We started with an initial sample of 52 LMXBs for which there were either Chandra or XMM-Newton observations. The majority of these systems (23 BHs and 23 NS respectively) are transient, low-to-moderate accreting sources. The sample also included two microquasars and four LMXBs with an uncertain nature. Of these we identified 40 sources with distance information so that we could establish a luminosity scale. As a preliminary step, we extracted spectra for all these systems in order to identify sources with reasonable signal-to-noise ratio and sufficient counts in the 0.5 - 10 keV energy band to allow for further detailed spectral analysis. Our final sample consists of 27 LMXBs (15 BH and 12 NS binaries respectively) for which we present a detailed spectral analysis. We note that in our final sample all BH systems and the majority of the NS systems (with a couple of exceptions) show transient behavior.  The entire data sets are taken from the Chandra and XMM-Newton archives. We direct the reader to Tables 1, 2, and 3 for full details regarding number of data sets analyzed, the particular observation IDs, and the particular models deployed in the spectral fits. Consistent with one of our starting criteria, the chosen targets fall into the requisite range in luminosity i.e., from a few x $\sim 10^{30}$ erg/s to $10^{36}$ erg/s. As designed, this extends the lower luminosity threshold considerably compared to \citet{2015MNRAS.454.1371W} who chose $10^{34}$ erg/s as the cutoff. Their study and others \citep{2001ApJ...551..921R, 2002ApJ...577..346R, 2004ApJ...610..933T, 2002ApJ...575L..15C, 2005ApJ...618..883W, 2007ApJ...660.1424H, 2009ApJ...691.1035H} hint at evidence that the situation regarding XRB spectra, especially the spectral behavior of certain groups of NS binaries becomes very complex at or around this threshold. In this study we have embarked on probing whether indeed there is some level of 'bifurcation' among these NS groups as a function of luminosity, and if so, whether these groups can be isolated for further study particularly in regards to the appearance of the SPL spectral feature which might be indicative of a new emission mechanism responsible for their apparent diverse behavior. \\ 
\\
For the Chandra data set, we used version 4.6 of CIAO with the most up-to-date calibration files to reduce the $evt2$ files. Using the $specextract$ script, source and background spectra and responses were created from each observation. The source spectra were extracted using a circle centered on the source. Background spectra were extracted from an adjacent, source-free, region of the same size. For the XMM measurements, observation data files (ODFs) were processed using standard procedures of the \emph{XMM-Newton} Science Analysis System (XMM-SAS version 13.0.1). For each observation, spectra were extracted using circular regions centered on the X-ray source of interest. Background subtraction was performed using source-free apertures located nearby or around the source. The HEASARC/FTOOLS suite was used to further process the data; the routine $grppha$ was used to group all spectra.\\
\\
The spectra were analyzed using XSPEC version 12.8.2 \citep{1996ASPC..101...17A}. The data were fitted over the 0.3 - 10.0 keV energy range: models considered include the PL (power-law) and thermal models, such as {\it diskbb} and {\it bbody}, along with galactic and intrinsic absorption. The galactic N$_H$ was fixed and the intrinsic N$_H$ was allowed to vary. As an initial step, we used a simple powerlaw (PL), and then added a soft component using either a standard blackbody (bbody) or an accretion disc blackbody model such as (diskbb). We did this to ascertain whether there were significant differences in the extracted parameters particularly the spectral indices. For the majority of our fits,  11 of 15 BH sources and 7 of 12 NS sources respectively, a PL by itself produced sufficiently acceptable fits with $\chi^{2}$ $\leqslant$ 1.3. These represent significant fractions of the overall samples for the two respective groups. For the remaining spectra  (4/15 BH and 5/12 NS), a PL by itself led to a $\chi^{2}$ $\geq$ 1.3 and required an additional spectral component.  Another method that is sometimes deployed to discern whether an additional model is needed to describe the spectrum is the F-statistic; in this method (which should be used with some caution since it is known to either fail or lead to inconclusive results under certain conditions; see \citet{2002ApJ...571..545P}), the fit results and the sum of the squares of the residuals obtained with a full model (typically consisting of a set of linear models), are compared with fits and residuals obtained with a restricted model (formally a subset of the full model). A comparison of the F-statistic obtained from the data and that calculated from the underlying F-distribution (for the relative change in the degrees of freedom in going from the full model to the  restricted model), provides a test for either accepting  or rejecting (at a certain confidence level) the hypothesis that an additional component is necessary in fitting the data. We performed this test for the NS binaries assuming the full model to consist of a PL plus a thermal component (either a $bbody$ or a $diskbb$). The restricted model was simply either the PL or one of the thermal components. The F-statistic was computed for all combinations for the NS sample: Six (out of 12), the ones with large spectral indices, passed the F-test, indicating that an additional component, beyond the PL, is not needed at the 95$\%$ confidence level. This result is consistent with our use of the $\chi^{2}$ cutoff. The F-test for 5 NS sources indicated the need for an additional component beyond the simple PL, and the results for one were inconclusive. The models used and the parameters extracted from the fits are summarized in Table 1. A glance at the fit parameters shows that for the 6 NS sources that exhibit the softest spectra (largest $\Gamma$), the PL was the only model used in fitting the data. An additional component did not produce any significant improvement in the $\chi^{2}$. As the combined results show (see later section), these are the sources that show the steepest slope as a function of luminosity. Secondly, we fitted both the Chandra and XMM data sets with the physically motivated Comptonization model, COMPPS. These fits allow us to assess the contribution of Comptonization, especially the role played by low-energy seed photons emanating from the accretion disk. In this portion of the analysis, we followed the strategy adopted by \citet{2017MNRAS.466..194B} who fixed several parameters including R (the amount of reflection; assumed 0.55 for NS and 0.24 for BH binaries). The electron distribution was assumed to be Maxwellian and the geometry was assumed to be spherical. Further, the PHABS model was used to account for the  absorption column, with N$_H$ fixed at the galactic level. Depending on the quality of the fit, we also fixed the blackbody temperature, and the normalization in some cases. The corona electron temperature and the optical depth were left as free parameters. The X-ray luminosities were computed in the range 0.5 - 10 keV. The key parameters extracted in this portion of the analysis, including the blackbody and corona temperatures, kT$_b$ and kT$_e$ respectively, and optical depths $\tau_y$, are listed in Table 2.\\

\begin{figure}
 \vspace{-0.2cm}  
\begin{minipage}[b]{1.0\linewidth}
\centering      
\includegraphics[scale=0.7, angle = 0]{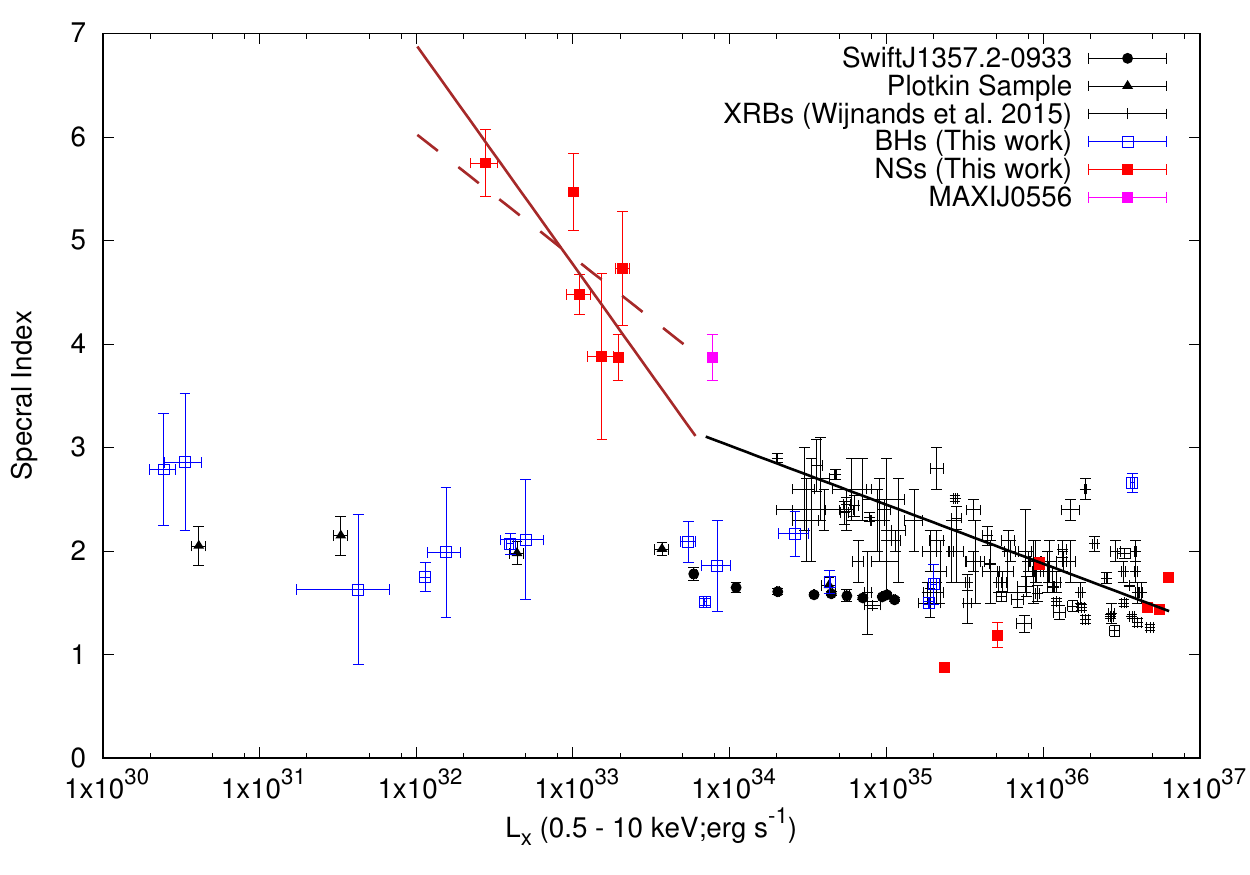}
\includegraphics[scale=0.7, angle = 0]{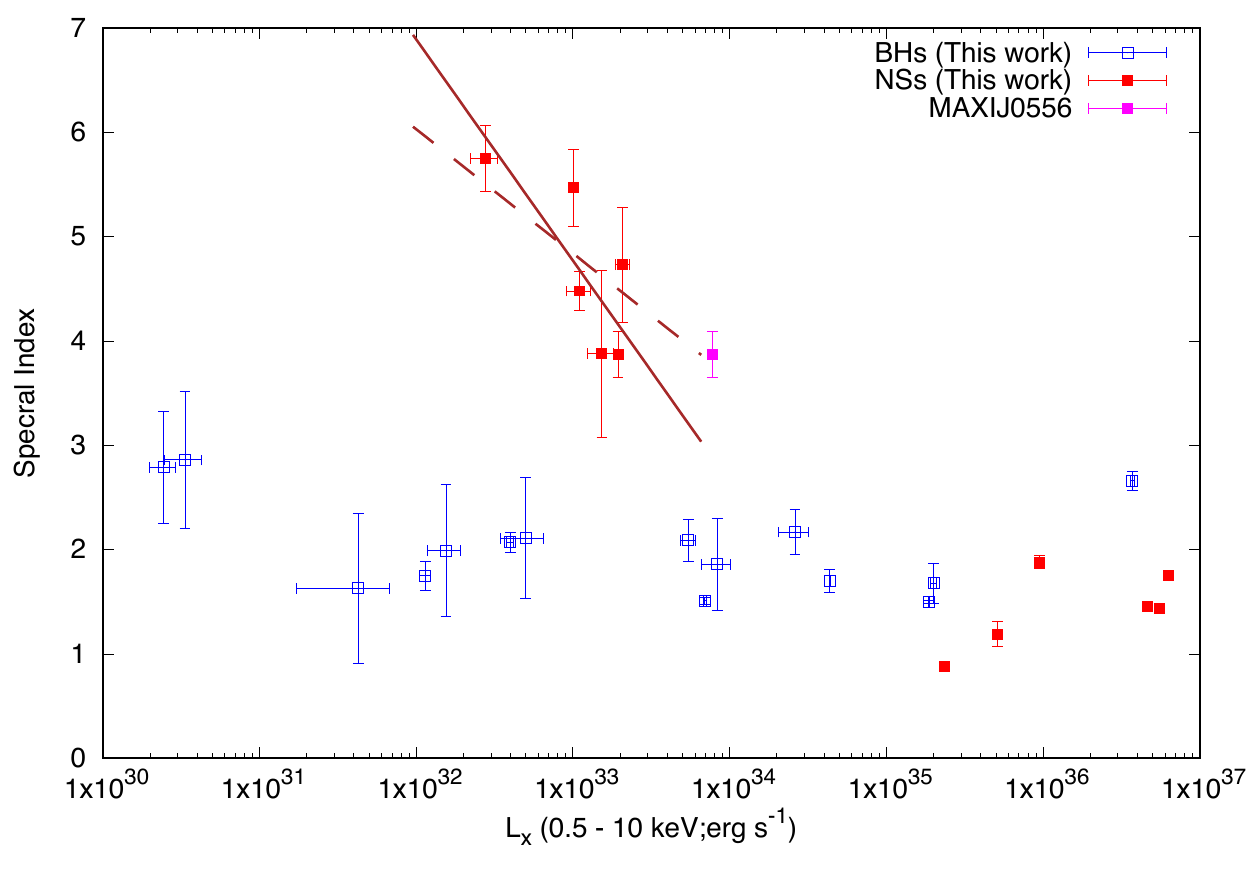}
\includegraphics[scale=0.7, angle = 0]{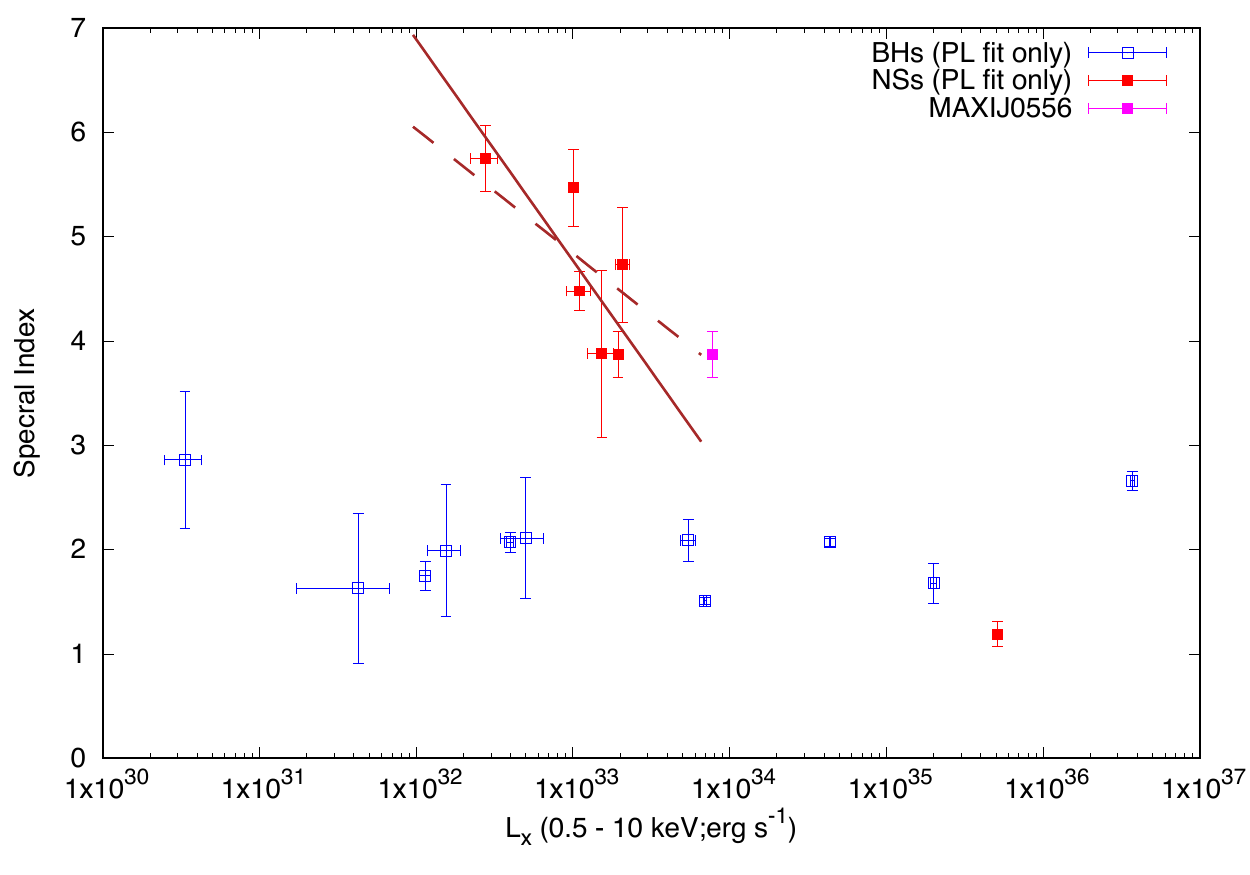}
\caption{(a) top panel: Spectral index vs. X-ray luminosity in the 0.5-10 keV for BH binaries (blue), and NS binaries (red). The results from the \citet{2015MNRAS.454.1371W} study are shown in black. The black line is a fit to the high-luminosity data. The red solid line is a fit to the low-luminosity NS data. The dash red line is same fit but with different luminosity for source MAXIJ0556-332. (b) middle panel: Same variables as above but for sources investigated in this work. (c) bottom panel: Our sample with PL fits only.}
    \vspace{-0.2cm} 
    \label{fig1}
\end{minipage}
\end{figure}
\begin{figure}
\centering      
\includegraphics[width=9cm, angle = 0]{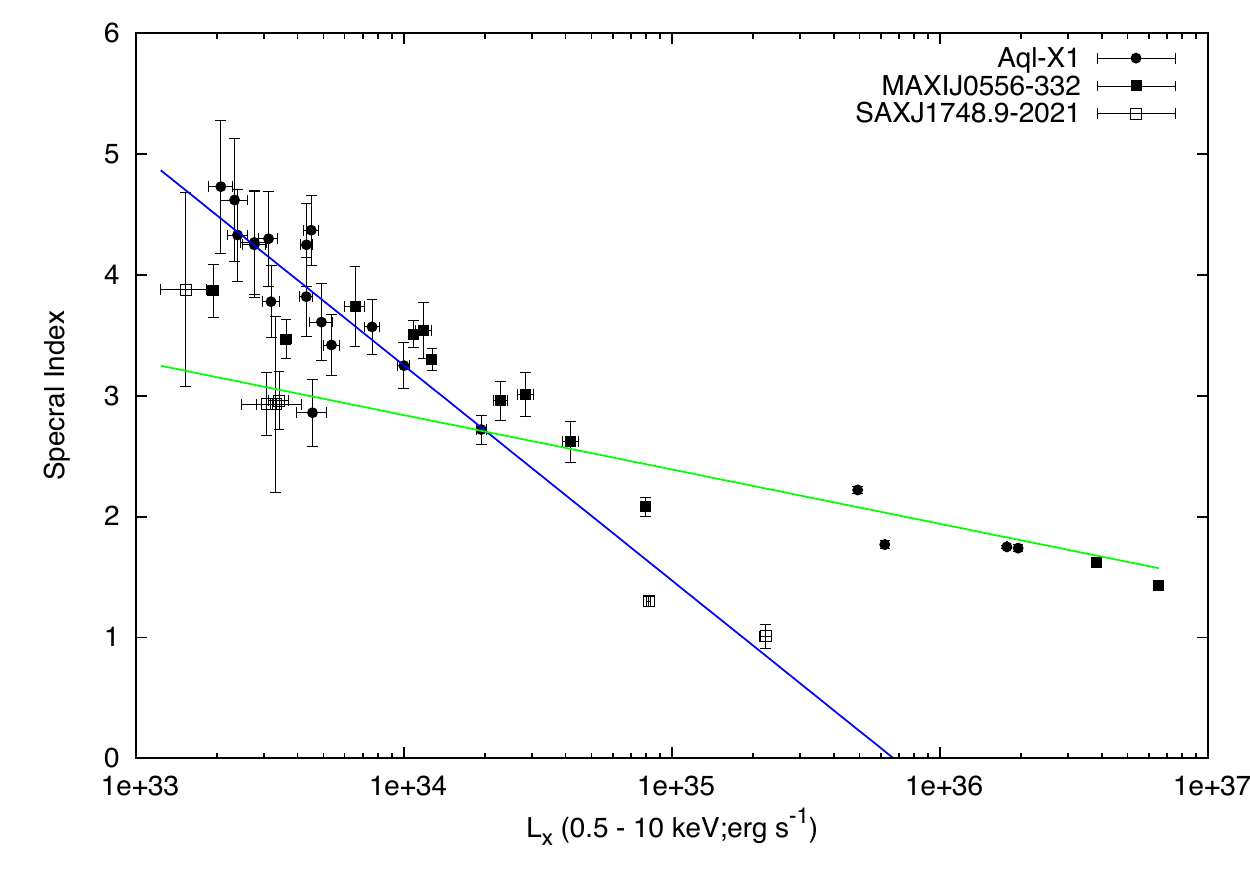}
\vspace{-1.0 cm}
\caption{\footnotesize{Spectral index vs. X-ray luminosity in the 0.5-10 keV for three (NS) LMXBs. The blue line is a fit to the low-luminosity data only. The green line is a fit to the high-luminosity data. An apparent transition of spectral behavior occurs at a luminosity of $\sim$ 10$^{35}$ erg/s.}}
\end{figure}
\begin{figure}
 \vspace{0.3cm}  
\begin{minipage}[b]{1.0\linewidth}
\centering      
\mbox{\includegraphics[scale=0.65, angle = 0]{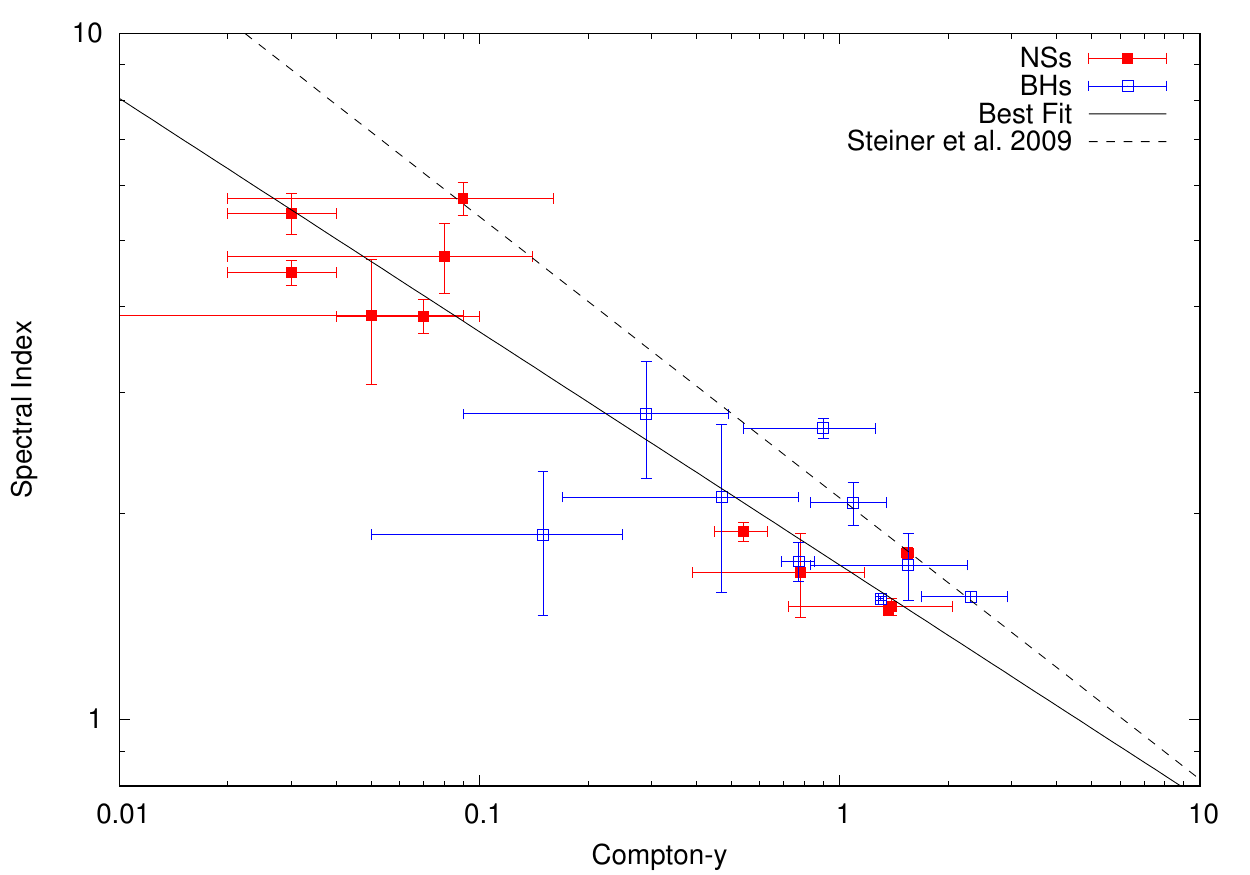}}   
\caption{Spectral index vs. the Compton-y parameter for BH binaries (blue), and NS binaries (red). Solid line is the best-fit to the data and the dashed line is a model calculation \citep{2009PASP..121.1279S}.}
    \vspace{-0.4cm} 
    \label{fig1}
\end{minipage}
\end{figure}

\begin{figure}
 \vspace{-0.2cm}  
\begin{minipage}[b]{1.0\linewidth}
\centering      
\mbox{\includegraphics[scale=0.45, angle = 0]{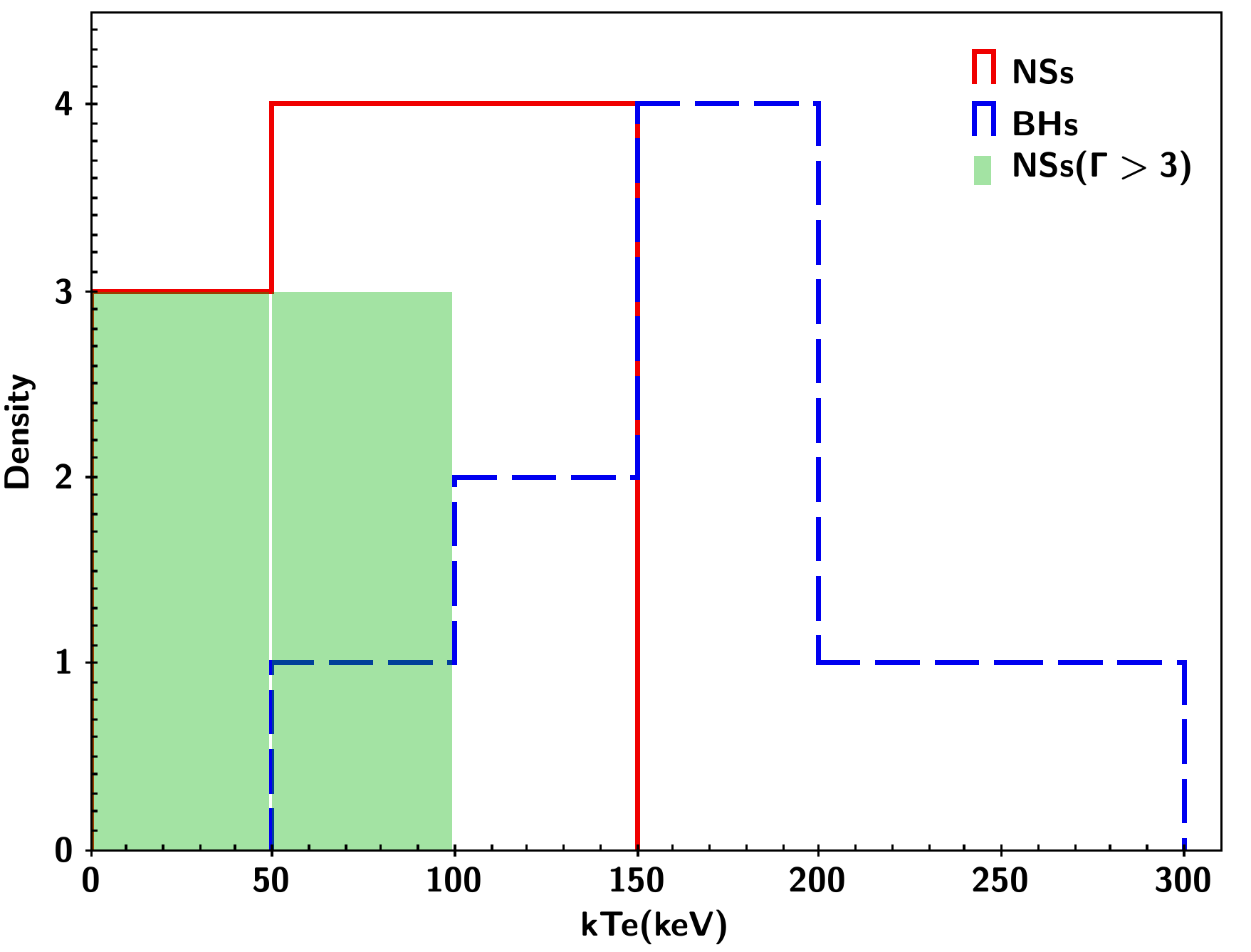}}   
\caption{Distribution of corona temperature (kTe) for BH binaries (blue), and NS binaries (red). Green block indicates NS sources with soft spectra i.e., $\Gamma$ $>$ 3.}
    \vspace{-0.4cm} 
    \label{fig1}
\end{minipage}
\end{figure}


\begin{figure}
 \vspace{-0.2cm}  
\begin{minipage}[b]{1.0\linewidth}
\centering      
\mbox{\includegraphics[scale=0.60, angle = 0]{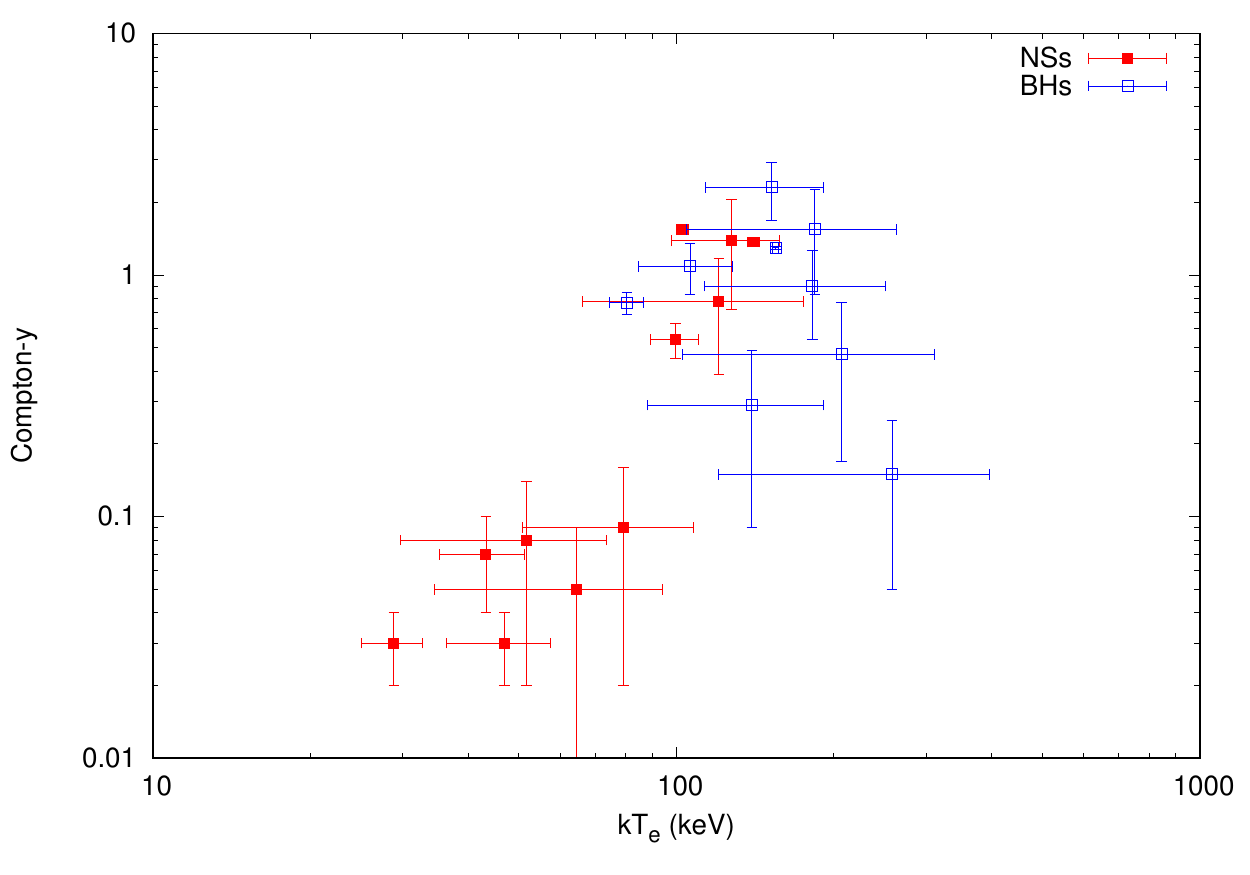}}   
\caption{Compton-y parameters vs kTe for BH binaries (blue), and NS binaries (red).}
    \vspace{-0.4cm} 
    \label{fig1}
\end{minipage}
\end{figure}



\begin{figure}
 \vspace{-0.1cm}  
\begin{minipage}[b]{1.0\linewidth}
\centering      
\mbox{\includegraphics[scale=0.70, angle = 0]{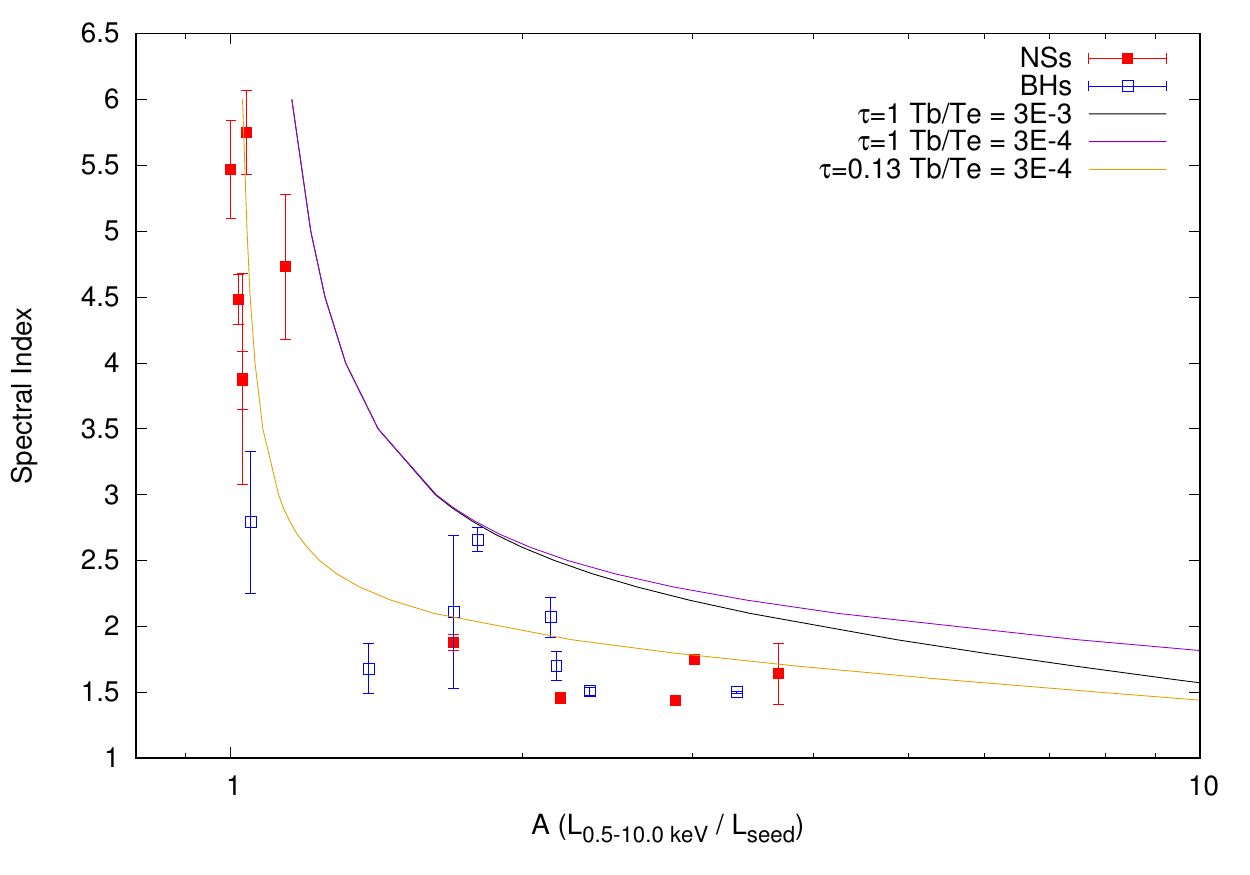}}   
\caption{Spectral index vs Amplification factor for BH binaries (blue), and NS binaries (red). The curves show results of calculations of Comptonization based on a disk-corona model \citep{016gilfanovm.pdf} for varying temperatures and optical depth.}
    \vspace{0.1cm} 
    \label{fig1}
\end{minipage}
\end{figure}

\section{Results and Discussion}
In Figure 1(a), we present the collective trend of the spectral index of our sample of LMXBs (BHs: blue, NS; red), combined with the sample of \citet{2015MNRAS.454.1371W} (black), with respect to the corresponding X-ray luminosity in the 0.5-10 keV range.  Our sample of 27 sources is shown by itself in Figure 1(b). On the whole, the results from our larger sample are in reasonable accord with the findings reported by \citet{2015MNRAS.454.1371W} in that the NS binaries exhibit significantly more softer spectra in the low-luminosity regime as compared to BH binaries over the same range. We emphasize here that the group of (6) NS binaries that show the softest spectra (large spectral indices i.e., $\sim$3 - 6) are well fitted with a PL only i.e., for these sources an admixture of models is not needed for a good spectral description in the chosen energy band (see Figure 1(c)). However, for several NS-sources (4U 1702-429, 4U 1728-16, 4U 1820-30 and 4U 1850-087) a thermal component was added to constrain the fit and achieve a reasonable $\chi^{2}$. Both sets of fit parameters (PL only and PL+ thermal component) are listed in Table 1. For 4U 1702-429, the PL fit was rejected because it produced an unacceptably large $\chi^{2}$. In all of these cases the extracted spectral index varies between $\sim$ 0.8 - 2.1 i.e, typical of relatively hard spectra and certainly very small compared to the indices for the group of 6 NS binaries that exhibit the steep slope as function of luminosity. For the NS sample as a whole, the spectral index rises more steeply in the low-luminosity range compared to that in the range above $\sim$10$^{35}$ erg/s. This is illustrated by the fits (red solid and dashed lines) to the NS-data (red points) where we have employed the function: $\Gamma$ = alogLx + b. We find a slope of $\sim$ -2.12 +/- 0.63 for the solid line. The dashed line is obtained when we use a different distance (Homan et al. 2014) to calculate the luminosity for the source: MAXIJ0556-332. The slope in this case is -1.19 +/- 0.34. These steep slopes in the low-luminosity region are to be compared with the slope of -0.42 +/- 0.04 obtained by \citet{2015MNRAS.454.1371W} for their NS-data in the higher luminosity range. If we combine our high-luminosity NS data with those of \citet{2015MNRAS.454.1371W}, we obtain a slightly larger slope (-0.57+/- 0.03) (shown as a black line in Figure 1(a)) but nonetheless the difference between the low-luminosity data (10$^{32}$  -  10$^{34}$ erg/s) and the high-luminosity (10$^{34}$  -  10$^{37}$ erg/s) is clearly significant. The error bars on some of the extracted spectral indices are relatively large as indeed was noted by \citet{2015MNRAS.454.1371W}. No doubt this is in part due to the lack of photon statistics for the low-luminosity data sets and the limited model parameters deployed in the fits (only a PL fit in majority of the cases). Another obvious consideration, as noted in the introduction, is related to the spectral state of the individual source. Following Wijnands et al (2015), we have focused on the lowest luminosity scale for each of the observations i.e., sources apparently in the LHS or that approaching the quiescent state, and where possible, the use of the simplest spectral model in fitting the data. Of course we are cognizant of the fact that the luminosity scale by itself is not a proxy for the designation of a spectral state especially if the state falls into the somewhat ambigous SPL state which encompasses a large region in the HID.\\ 
\\
Based on the low-luminosity region of Figures 1(a,b), the overall trend is clear in that the spectral index for BH binaries tends to plateau at low luminosities whereas the index for the NS binaries is anticorrelated with the luminosity and displays a steep increase as the luminosity decreases indicating much softer spectra in the range of luminosities that might typically be associated with the LHS. This anticorrelation has been noticed before: see \citep{2004MNRAS.349...94J,  2004MNRAS.354..666J}. These authors note a correlation between the fractional power-law contribution and the (0.5 - 10 keV)- luminosity above $\sim$10$^{33}$ erg/s and an anticorrelation between these two quantities for luminosities below $\sim$10$^{33}$ erg/s. While our data do not explicitly exhibit both features, the data do however suggest a transition i.e. a marked change of slope in the spectral-index-luminosity trend around a luminosity of $\sim$10$^{34}$ erg/s. \\
 \\
For a few of our NS binaries multiple observations exist; we analyzed all these observations for Aql-X1, MAXIJ0556-332, and SAXJ1748.9-2021 in order to get a sense of their individual behavior as function of luminosity i.e., whether they individually exhibit the global feature observed in the NS sample as a whole. We fitted the available spectra with a phenomenological PL and extracted the spectral index ($\Gamma$). In Figure 2, we show the extracted photon index ($\Gamma$) as a function of X-ray luminosity in the 0.5-10 keV range. We find that the sources exhibit significantly soft spectra in the low-luminosity regime i.e., $\Gamma$ rises steeply in the low-luminosity range compared to that in the range above $\sim 10^{35}$~erg/s. This is illustrated by the separate fits to the low- and high-luminosity data, where we have employed the same function as Figure 1a. We find a slope of -1.78 $\pm$ 0.14 in the low-luminosity region (indicated by the blue line fit in Figure 2), and a slope of -0.45 $\pm$ 0.17 in the high-luminosity region (green line fit);  the trend of the low-luminosity data and the high- luminosity data is significantly different. Admittedly the data are limited and the transition is at best convincing for only two of the three sources (namely Aql-X1 and MAXIJ0556-332) but nonetheless the data do suggest, at least in a limited number of individual sources, that the softening feature appears to evolve as a function of luminosity and is likely to be a general feature in a larger set of similar low-accreting systems.\\
 \\         
In addition to our primary aim, that of probing the $\Gamma$ - Luminosity correlation, we also wanted to explore whether the data could be described by a physically motivated model such as Comptonization for it has the necessary ingredients for the production of thermal and non-thermal components in a self consistent manner. In a recent study, \citet{2016MNRAS.458.1778A}, using frequency-resolved spectra suggest further evidence for two major Comptonization regions in BH binaries i.e., a soft one, the origin of which is associated with the disk, and a hard one which is apparently independent of the disk. They further suggest that both are likely to be inhomogeneous. While we are not in a position to perform a similar study, we are nonetheless motivated to explore whether the presence of these components, particularly the soft one, can be identified through a Comptonization model such as COMPPS (in XSPEC). We used the COMPPS model to extract the Compton-y parameter for our data sets. The results are shown in Figure 3, where we have plotted the spectral indices extracted via PL fits of the same data sets. Despite the fairly large uncertainties, the Compton-y parameters and the photon indices ($\Gamma$) clearly exhibit an anticorrelation. The expected theoretical dependence on $\Gamma$ is indicated by the dashed line and is based on a solution to the Kompaneets equation (see \citet{1976ApJ...204..187S, 1979AstQ....3..199R}). For completeness, also shown is the best fit to the data, solid line with an exponent of -0.34+/- 0.05 which is somewhat shallower than the result (-0.41 +/- 0.01) from the physically motivated model of Comptonization via repeated scattering by non-relativistic electrons. In addition, notice that the low Compton-y parameter region (corresponding to low optical depth), is solely occupied by (6) NS binaries (red points) that have the softest spectra i.e, the ones with largest photon index (see Fig. 1). In contrast, all the BH binaries occupy the mid-to-high values of the Compton-y parameter region (corresponding to relatively high optical depth). These sources exhibit relatively hard spectra. We also notice that a number of NS binaries occupy this high end of the Compton-y parameter region as well but we point out that this group is about an order of magnitude above the low-luminosity threshold of $\sim$10$^{34}$ erg/s. In addition to the Comp-y parameter, the electron cloud (corona) temperature, kTe (keV), is another key parameter of the Comptonization model: a histogram of the extracted kTe is plotted in Figure 4. The NS distribution (shown in red), with a mean temperature around 90 keV,  is clearly displaced from the BH distribution (shown in blue), with a mean value of approximately 175 keV. This significant offset between the two groups of binaries again suggests a spectral signature for distinguishing between the two groups. This is in agreement with the findings of \citet{2017MNRAS.466..194B} who also report a clear separation of kTe scales for the BH and NS groups. Shown as shaded green is the group of 6 NS binaries with the relatively soft spectra; the group is clustered at a significantly lower corona temperature ($\sim$50 keV) than the average temperature for the NS sample ($\sim$90 keV).
The expected correlation between Comp-y and the corona temperature kT$_{e}$ is nicely illustrated in Figure 5. The separation between the group of soft-spectra 6-NS binaries and the rest of the sample is clearly visible. \\
\\
The impact of the seed photons on the Comptonization process is quantified by the blackbody temperature (kT$_{bb}$): The majority of the NS binaries are well within the typical disk temperatures of about 0.1 - 0.2 keV. A large fraction of the BH sources also lie in the same range although the dispersion is somewhat larger than the NS sample. The 6 soft-spectra NS binaries reside comfortably in the low-temperature range. In addition, we follow the procedure defined by \citep{2017MNRAS.466..194B} and calculate the Compton amplification factor (A) that provides a measure of the relative contributions to the thermal and non-thermal components of the emitted spectrum (essentially a proxy for the real factor because of the limited energy band of our data sets) for all the sources fitted with the COMPPS model. A simplified configuration envisages the emission of power in a hot optically thin region in the vicinity of a much cooler optically thick source. The two sources (hot corona and the cool disk) are understood to be coupled in that the emission from the cool source is the origin of the soft photons for the Comptonization process and that the more energetic or harder photons that are produced as result of Comptonization, in turn, produce heating of the disk. This feedback, by way of an energy balance, then determines the Comptonized amplification factor A. Of course since the feedback determines the relative fractions of the emitted power that leads to the emitted spectrum consisting of the three components i.e., the thermalized component from the thick disk (blackbody-like), the power-law from direct Comptonization, and the reflection component, the amplification factor is thus expected to be intricately connected to the emitted spectrum. In Figure 6, we show the behavior of the extracted spectral indices for our sample as a function of the Comptonization amplification factor (normalized to unity primarily because of the very limited energy-band coverage of our data sets). Along with the data, we also show the results of model calculations for the relation between the spectral index and the amplification factor based on simple disk-corona model \citep{016gilfanovm.pdf}. We see that the data closely follow the predicted theoretical trend for parameters pertinent to our data (green curve) i.e., optical depth $\sim$0.13 and (Tbb/Te) $\sim$10$^{-4}$. Moreover, we notice that the NS binaries with the largest spectral indices appear to be asymptotically approaching the theoretical limit of A$\sim$1. i.e., the spectra are considerably softer compared to the comparable luminosity BH binaries which are also approaching the same limit but relatively slowly i.e, BH binaries have significantly harder spectra for the same amplification factor.\\
\\
Assuming a hot corona with mildly relativistic bulk motion, \citet{1999ApJ...510L.123B} suggests a different relation between the spectral index and the amplification factor i.e., $\Gamma$$\sim$ A$^{-k}$, where the exponent is estimated to be 1/6 for galactic BHs. Using this relation, a fit to our BH data leads to an exponent of 0.35 +/- 0.09. This value seems to be in tension with the model prediction of 1/6 for BHs thus arguing against the presence of relativistic bulk motion. For the NS data we find a similar exponent i.e., 0.27 +/- 0.03, suggesting insensitivity to the nature of the compact object. However, we recognize that our sample is small and the overall statistics are relatively poor thus our statements are necessarily speculative at best. Moreover, since we are mostly interested in the comparison of what seems to be a universal behavior of LMXBs in the low-luminosity regime, we are less concerned with the absolute value of the amplification factor and more with its overall trend as a function of the spectral index, $\Gamma$.\\
\\
Wijnands et al. 2015 place the luminosity threshold for NS binaries at 10$^{34}$ erg/s: it is noted that below this threshold the spectral situation becomes complex i.e., some systems are totally dominated by the thermal component leading to very large spectral indices \citep{2001ApJ...551..921R, 2002ApJ...577..346R, 2004ApJ...610..933T}, while other systems tend to exhibit significant PL contribution with fairly low spectral indices \citep{2002ApJ...575L..15C, 2005ApJ...618..883W, 2009ApJ...691.1035H, 2012ApJ...756..148D}. Our study confirms the large spectral indices (in the range 3 - 6) for a group of (6) NS binaries. It is certainly plausible (as suggested by \citet{2015MNRAS.454.1371W}) that the observed large thermal component arises as a result of the NS surface i.e., cooling of the surface as the NS systems transition from outburst to and through the quiescent state. Indeed, some recent studies \citep{2004ApJ...606L..61W, 2008ApJ...687L..87C, 2015MNRAS.451.2071D, 2011ApJ...736..162F, 2011A&A...528A.150D, 2014ApJ...795..131H, 2016MNRAS.456.4001W, 2016ApJ...833..186M} that have monitored a sample of transient LMXBs in their quiescent states have been used to extract cooling curves that allow the determination of effective NS surface temperature profiles. Although not directly comparable, the effective seed photon temperatures extracted in our study for the group of NS binaries with high spectral indices (see kT$_{bb}$ column in Table 2), fall in a range (80-170 eV) that is not too dissimilar from the effective NS surface temperatures extracted by \citet{2014ApJ...795..131H} in their study of a number of transient NS binaries in the quiescent state. This is unlikely to be a coincidence. Clearly additional data are needed to put this on a more firm footing but the similarity of the extracted temperature ranges is suggestive of the notion that the NS surface likely plays a significant role in the thermal emission observed in a number of NS binaries in or near their quiescent state.
\section{Summary}
We have performed a spectral analysis of a sample of LMXBs in the low-luminosity regime. The luminosity coverage is large, ranging from $\sim$10$^{30}$ to $\sim$10$^{36}$ erg/s. Our sample includes persistent and transient BH and NS binaries; AMXPs are excluded. The analysis has been performed in two stages: in the initial stage the spectra were primarily fitted with a phenomenological PL for a relatively large sample of NS and BH binaries across a wide range of luminosities. For a few cases a thermal component was added in order to achieve an acceptable $\chi^{2}$. The adopted strategy was not to necessarily strive for the best possible statistical fits but to keep the fit parameters to a minimum and a common set so that global trends among the data sets, if any, could be robustly compared. In this part of the analysis we followed the procedure adopted by \citet{2015MNRAS.454.1371W}, who analyzed a smaller sample of sources and within a more limited luminosity range and presented evidence for a dichotomy of behavior of NS and BH binaries, cast as a correlation between the spectral index and the luminosity. In the second part of the analysis, we followed the strategy of \citet{2016MNRAS.tmp.1505B}, by deploying the more physically motivated model of thermal and bulk Comptonization to explore the BH-NS dichotomy in detail. \\
\\
We summarize our main findings as follows:
\begin{itemize}
\item We find evidence for a significant anticorrelation between the spectral index and the luminosity for a group of NS binaries in the luminosity range  $\sim$10$^{32}$ to $\sim$10$^{33}$ erg/s. Our analysis suggests a steep slope for the correlation i.e., -2.12 +/- 0.63. At the higher luminosities, we find a shallower slope in agreement with the findings of \citet{2015MNRAS.454.1371W}. In contrast, BH binaries do not exhibit the same behavior. 
\item  The NS-BH dichotomy is further demonstrated by the corona temperature (kT$_{e}$) extracted for the bulk of the sources; the mean temperature for the NS group is significantly lower than the equivalent temperature for the BH group, in good agreement with the findings of \citet{2017MNRAS.466..194B}. 
\item We show that the extracted Comptonized amplification factor, A, follows the  theoretically predicted relation with the spectral index, $\Gamma$. This lends support to a basic Comptonization model which assumes a feedback mechanism between an optically  thin hot corona and an optically thick cooler source of soft photons. The low-luminosity NS binaries tend to approach the amplification limit more rapidly compared to the BH binaries.  
\item A formulation of the Comptonization process in terms of a hot corona with mild relativistic bulk motion predicts a simple power-law relation between $\Gamma$ and the amplification factor, A. For reasonable parameters, this model predicts an exponent of 1/6 for galactic BHs; fits to our BH data suggest an exponent of 0.35 +/- 0.09 i.e., in tension with the model prediction and arguing against the presence of relativistic bulk flow. Fits to the NS data produce a similar exponent, suggesting insensitivity to the nature of the compact object. We note however the precision of the current data is insufficient to justify a more definitive conclusion.
\item The photon seed temperatures extracted for a number of NS binaries with relatively large spectral indices fall in a range that is not too dissimilar from the effective NS surface temperatures extracted from cooling curves of transient NS systems in their quiescent state. If verified by additional data, this would lend support to the idea that the NS surface plays a significant role in the observed thermal emission from these systems in or near their quiescent state.
\end{itemize}

We would like to acknowledge the kind support of TUBITAK-BIDEB 2221 and the Sabanci University Faculty of Engineering and Natural Sciences where part of this work was completed during KSD's sabbatical stay. We thank the Referee for constructive comments. Finally, we would like to take this opportunity to pay tribute to our friend and colleague, Neil Gehrels, who was an inspiration to all of us.\\

\pagestyle{empty}
\clearpage
\begin{longrotatetable}
\begin{deluxetable}{ccccccccc}
\tablecaption{Summary of fit parameters for our sample of LMXBs in the Low-Luminosity range.}
\tabletypesize{\footnotesize}
\tablewidth{0.99\textwidth}
\tablehead{
\colhead{Source Name} & \colhead{Observation ID} & \colhead{Distance} & \colhead{Luminosity (0.5 - 10 keV)} & \colhead{Model}& \colhead{$\Gamma$} & \colhead{Tin} & \colhead{kT$_{bb}$} & \colhead{$\chi^2$/dof}\\
\colhead{} & \colhead{(X)-XMM; (C)-Chandra} & \colhead{kpc} & \colhead{erg~s$^{-1}$} &\colhead{} &\colhead{} & \colhead{keV}& \colhead{keV} & \colhead{}\\
} 
\startdata \\
		{\bf BH Binaries} \\
		4U 1543-47&0155762201(X) & 9.1   & 4.35$\pm$0.05$\times10^{34}$& PL+diskbb & 1.7$\pm$0.1 & 0.70$\pm$ 0.04 & \nodata & 2317.1/2400\\
				  &             &      &      &  PL& 2.07$\pm$0.04   &    &      & 2334.1/2402\\
		A 0620-00   &95(C) & 0.87  & 2.5$\pm$0.5$\times10^{30}$ & PL& 2.8$\pm$0.5 &   \nodata & \nodata  & 34.4/25\\
		GRO J1655-40& 0400890301(X)& 3.2 & 1.14$\pm$0.09$\times10^{31}$ & PL& 1.8$\pm$0.1 & \nodata & \nodata &90.1/91\\
		GS 1124-68& 0085960101(X) & 5.5 &4$\pm$3$\times10^{31}$ & PL&  1.6$\pm$0.7 &\nodata  & \nodata &66.3/132\\
		GS 1354-64 & 15576(C)&  25  &8$\pm$2$\times10^{33}$  & PL& 1.9$\pm$0.4 &\nodata &\nodata & 29.3/22\\
		GX 339-4 & 0085680501(X)&  10  &5.5$\pm$0.6$\times10^{33}$ & PL& 2.1$\pm$0.2 & \nodata &\nodata& 41.6/68\\
		MAXI J1659-152 & 12439(C)& 7 &5$\pm$2$\times10^{32}$ & PL& 2.1$\pm$0.6 &   \nodata &\nodata & 15.3/14\\
		V404 Cyg & 0304000201(X)& 2.39  &4.0$\pm$0.2$\times10^{32}$  & PL & 2.1$\pm$0.1& \nodata&\nodata & 396.8/408\\
		V4641SGR&  4451(C)&7.4  &2.6$\pm$0.6$\times10^{34}$ & PL+GAUSS& 2.2$\pm$0.2&\nodata & \nodata&52.5/50\\
	         XTE J1118+480& 14630(C)& 1.4 & 3.4$\pm$0.9$\times10^{30}$ & PL & 2.9$\pm$0.7& \nodata& \nodata&9.3/23\\
		XTE J1550-564 &  3672(C)&5.3 & 1.6$\pm$0.4$\times10^{31}$ & PL & 2.0$\pm$0.6&\nodata &\nodata &21.1/28\\
		XTE J1650-500 &2731(C) & 2.6 & 6.9$\pm$0.2$\times10^{33}$ & PL &1.51$\pm$0.03 &   \nodata&\nodata& 274.7/283\\
		1E 1740.7-2942& 658(C)& 8.5 & 2.0$\pm$0.1$\times10^{35}$ &PL &1.7$\pm$0.2&  \nodata &\nodata & 210.9/238\\
		GRS 1758-258 & 0112971301(X) & 8.5 & 3.7$\pm$0.1$\times10^{36}$ & PL& 2.66$\pm$0.09&   \nodata& \nodata& 341.1/403\\
		Swift J1357.2-0933& 00674580101(X) &2.3 - 6.3 & 1.87$\pm$0.01$\times10^{35}$  & PL+diskbb& 1.50$\pm$0.01&0.16$\pm$0.01 &\nodata&  2332.9/1796\\
		{\bf NS Binaries} \\
		4U 1608-52 & 12470(C) &  3.3 &1.1$\pm$0.2$\times10^{32}$ & PL & 4.5$\pm$0.2&   \nodata&\nodata& 52.3/47\\
		4U 1908+005 (Aql X1) &  3489(C)& 5.2 & 2.1$\pm$0.2$\times10^{33}$ & PL& 4.7$\pm$0.6 & \nodata&\nodata &26.5/29 \\
		MAXI J0556-332 & 0744870201(X)& 17 & 1.94$\pm$0.09$\times10^{33}$ & PL & 3.87$\pm$0.09 &   \nodata&\nodata& 246/249\\
		SAX J1748.9-2021& 0149180901(X)& 8.5 & 1.5$\pm$0.3$\times10^{33}$ & PL & 3.9$\pm$0.8 &\nodata &\nodata &730.3/1028\\
		SAX J1750.8-2900&  14651(C) &6.79 &2.8$\pm$0.5$\times10^{32}$  & PL& 5.8$\pm$0.3& \nodata&\nodata& 19.4/20 \\
		XTEJ1701-462& 0413390101(X)& 8.8 & 1.01$\pm$0.03$\times10^{33}$ &  PL & 5.5$\pm$0.4 & \nodata & \nodata &  179.5/176\\  
		4U 1702-429&0604030101(X)  &6.2 & 6.32$\pm$0.05$\times10^{36}$ & PL+bbody & 1.75$\pm$0.03 &   \nodata&4.0$\pm$0.2 & 7232.89/3855\\
		4U 1728-16 & 0090340101(X)& 4.4 &5.6$\pm$0.4$\times10^{36}$  & PL+bbody& 1.44$\pm$0.01 & \nodata&0.87$\pm$0.02 &4615.5/3063\\
				  &             &      &      &  PL& 1.47$\pm$0.01  &    &      &5478.3/3065\\
		4U 1728-34& 0701190101(X)& 5.3 & 4.63$\pm$0.06$\times10^{36}$ & PL& 1.46$\pm$0.04 &\nodata & \nodata & 2073.1/1544\\
		4U 1811-17& 0122340101(X)& 17.0 &5.1$\pm$0.2$\times10^{35}$ & PL &1.2$\pm$0.1& \nodata& \nodata& 1195.3/1158\\
		4U 1820-30& 0084110201(X)&7.6 & 2.34$\pm$0.08$\times10^{35}$ & PL+gauss&0.88$\pm$0.04& \nodata& \nodata& 1554.3/1621\\
		4U 1850-087& 0154150501(X)& 8.2 &9.51$\pm$0.03$\times10^{35}$ & PL+diskbb &1.88$\pm$0.06 & 0.58$\pm$0.01 & \nodata &2804.5/2071  \\
		&             &      &      &  PL& 2.18$\pm$0.01  &    &      &2925.9/2073\\
\enddata
\end{deluxetable}
\end{longrotatetable}
\clearpage

\pagestyle{empty}
\clearpage
\begin{deluxetable*}{ccccccc}
\tablecaption{Summary of COMPPS fit parameters for our sample of LMXBs in the Low-Luminosity range.}
\tabletypesize{\footnotesize}
\tablewidth{0.50\textwidth}
\tablehead{
\colhead{Source Name} & \colhead{Observation ID} & \colhead{kT$_{bb}$} & \colhead{kT$_e$} & \colhead{$\tau_y$} & \colhead{Amplification Factor} & \colhead{$\chi^2$/dof}\\
\colhead{} & \colhead{(X)-XMM; (C)-Chandra} & \colhead{keV} & \colhead{keV} &\colhead{} &\colhead{} &\colhead{}\\
} 
\startdata \\
		{\bf BH Binaries} \\
		4U 1543-47&0155762201(X) & 0.1 & 81$_{-6}^{+6}$ & 1.22$_{-0.07}^{+0.08}$ & 1.17 & 2459.5/2403\\
		A 0620-00   &95(C) & 0.14$\pm$0.02  &140$_{-45}^{+58}$ & 0.3$_{-0.1}^{+0.2}$ & 0.05 &32.1/26\\
		GS 1354-64 & 15576(C)& 0.11 &258$_{-100}^{+176}$ & 0.08$_{-0.02}^{+0.04}$ &0.18 & 25.6/22\\
		MAXI J1659-152 & 12439(C)& 0.1 &207$_{-80}^{+129}$ & 0.29$_{-0.09}^{+0.14}$ & 0.70 & 17.7/14\\
		V404 Cyg & 0304000201(X)& 0.35 & 106$_{-21}^{+23}$ & 1.3$_{-0.2}^{+0.2}$ & 1.14& 405.7/412\\
		XTE J1650-500 &2731(C) & 0.35 & 152$_{-35}^{+43}$ & 1.9$_{-0.2}^{+0.2}$ & 1.35& 256.0/282\\
		1E 1740.7-2942& 658(C)& 0.57$\pm$0.05& 184$_{-80}^{+80}$ & 1.1$_{-0.2}^{+0.2}$ & 0.39 & 206.3/237\\
		GRS 1758-258 & 0112971301(X) & 0.46$\pm$0.01& 182$_{-54}^{+83}$ & 0.63$_{-0.06}^{+0.09}$ & 0.80 & 319.1/401\\
		Swift J1357.2-0933& 00674580101(X) &0.12$\pm$0.01& 155$_{-2}^{+2}$ & 1.08$_{-0.01}^{+0.01}$ & 2.33 & 3165.50/1796\\
		{\bf NS Binaries} \\
		4U 1608-52 & 12470(C) &  0.15& 47$_{-10}^{+11}$ &  0.09$_{-0.03}^{+0.04}$ & 0.02 &54.9/45\\
		4U 1908+005 (Aql X1) &  3489(C)& 0.17 &52$\pm$22 & 0.2$\pm$0.1&0.14 &24.5/29 \\
		MAXI J0556-332 & 0744870201(X)& 0.08$\pm$0.01&43$_{-9}^{+8}$ &  0.22$_{- 0.07}^{+0.12}$ & 0.03& 242.3/247 \\
		SAX J1748.9-2021& 0149180901(X)& 0.11 & 64$_{-24}^{+35}$& 0.10$_{-0.05 }^{+0.09}$ & 0.03 & 746.4/1026\\
		SAX J1750.8-2900&  14651(C) &0.10& 79$_{-25}^{+32}$ &  0.15$_{-0.06}^{+0.12}$ & 0.04 & 27.3/20 \\
		XTEJ1701-462& 0413390101(X)& 0.12 & 29$_{-4}^{+4}$ &  0.14$_{-0.03}^{+0.05}$ & 0.002 &177.4/176\\  
		4U 1702-429&0604030101(X)  &0.10 &103$_{-3}^{+3}$ &  1.91$_{-0.05}^{+0.05}$ & 2.01 & 7876.8/3858 \\
		4U 1728-16& 0090340101(X)&  0.13$\pm$0.01 & 140$_{-4}^{+4}$ &  1.25$_{-0.01}^{+0.01}$ & 1.88   &4896.6/3068\\
		4U 1728-34& 0701190101(X)&  0.20 & 128$_{-33}^{+27}$ &  1.4$_{-0.4}^{+0.7}$ &1.19 & 1696.2/1569\\
		4U 1850-087& 0154150501(X)& 0.07$\pm$0.01 & 100$_{-10}^{+11}$ &  0.69$_{-0.08}^{+0.10}$ & 0.70 & 2709.1/2073 \\
\enddata
\end{deluxetable*}
\clearpage

\end{document}